\documentclass[
    ,final            
  ]
  {aipproc}

\layoutstyle{8x11single}

\usepackage{amssymb,amsmath}

\newcommand{\beq}{\begin{equation}}
\newcommand{\eeq}{\end{equation}}
\newcommand{\bitem}{\begin{itemize}}
\newcommand{\eitem}{\end{itemize}}
\newcommand{\bitemstep}{\begin{itemstep}}
\newcommand{\eitemstep}{\end{itemstep}}
\newcommand{\bcn}{\begin{center}}
\newcommand{\ecn}{\end{center}}

\begin{document}

\title{Explore Physics Beyond the Standard Model with GLAST}


\author{A.~M.~Lionetto}{
  address={INFN and Department of Physics Roma Tor Vergata\\
  Via della Ricerca Scientifica, 1\\ Italy.}
}

\begin{abstract}
We give an overview of the possibility of GLAST to explore theories beyond the Standard Model of particle physics. Among the wide taxonomy we will focus in particular on low scale supersymmetry and theories with extra space-time dimensions. These theories give a suitable dark matter candidate whose interactions and composition can be studied using a gamma ray probe. We show the possibility of GLAST to disentangle such exotic signals from a standard production background.
\end{abstract}

\maketitle

\section{Introduction}
\label{intro}
Dark matter still remains one of the main unsolved problem in
physics. The common accepted paradigm is the existence of an exotic
weakly interacting massive particle (WIMP). Such a particle has to be
found in some extension of the Standard Model (SM) of particle
physics. 
Supersymmetry and extra space-time dimensions are commonly used ingredients for a consistent theory beyond the SM.
One the most studied framework
is the MSSM, the minimal supersymmetric extension of the SM. In the
MSSM the lightest supersymmetric particle (LSP) is usually a
neutralino, which is a good candidate for cold dark matter (for a
recent review see~\cite{Bertone:2004pz}).
The pattern of the soft supersymmetry breaking terms and the presence of extra space-time dimensions greatly affects
the composition and the strength of the dominant interactions of the
neutralino.  
Hence it is crucial to study the neutralino phenomenology in different scenarios.
The GLAST experiment will have a great chance to shed light on the
nature of dark matter, for example through the analysis of the continuum
$\gamma$-ray flux supposed to come from pair WIMP annihilations.

\section{Exploring the CMSSM with GLAST\label{cmssm}}
One of the most studied supersymmetric scenario is the constrained MSSM
(CMSSM) in which the soft supersymmetry breaking terms are supposed to
derive from a high energy supergravity theory with a common scalar
mass $m_0$ and a common gaugino mass $m_{1/2}$ at the GUT scale.
In this framework the neutralino is the LSP in most of the parameter space. It can be studied through
indirect detection of $\gamma$-rays from pair annihilation. The
expected $\gamma$-ray continuum flux at a given photon energy $E$ from a
direction that form an angle $\psi$ is given by
\begin{equation} 
  \phi_\chi(E,\psi)=\frac{\sigma v}{4\pi} \sum_f \frac{dN_f}{dE} B_f
  \int_{l.o.s} dl(\psi) \frac{1}{2}\frac{\rho(l)^2}{m_\chi^2} 
\label{gammafluxcont}
\end{equation}
It depends on the particle physics model assumed
through the neutralino mass $m_\chi$, the total annihilation cross
section $\sigma v$ and through the sum of all the photon yield
$dN_f/dE$ per each annihilation channel weighted by
the corresponding branching ratio $B_f$. 
The flux~(\ref{gammafluxcont}) also depends on the WIMP density in the
galactic halo $\rho(l)$. The integral has to be performed along the
line of sight (l.o.s.). 
The WIMP density profiles $\rho(l)$ are in general
extremely cuspy towards the galactic center (GC) that hence is a good
place where to look for an exotic signal. In our analysis we take into account
the Navarro, Frenck and White (NFW) profile~\cite{Navarro:1995iw} and the Moore profile~\cite{Moore:1999gc}.   
The usual parametrization for a dark matter halo density is
\begin{equation}
  \rho(r)= \frac{\rho_0}{(r/R)^{\gamma}
  [1+(r/R)^{\alpha}]^{(\beta-\gamma)/\alpha}} \;\;.
\label{profile} 
\end{equation}
The NFW profile behaves like $r^{-1}$ towards the GC while the Moore
profile goes as $r^{-1.5}$.
In general one expects also contributions coming from standard astrophysical sources. 
In many diffuse continuum $\gamma$-ray production
models~\cite{Stecker:1977ua} in our galaxy, the dominant astrophysical
contribution comes from $\pi^0\to 2\gamma$. This has to be considered
the "standard" background. We performed a statistical analysis, based on
the usual $\chi^2$ test, in order to determine what are the models
that can be disentangled against the background by GLAST at a given
significance~\cite{Cesarini:2003nr}. We have taken into account
only the statistical errors, i.e. $\sigma=\sqrt{N_b}$ where $N_b$ is the
number of background photons in each energy bin.
   
Doing a detailed scan of the CMSSM parameter space defined by
$m_0$, $m_{1/2}$, $\tan\beta$, $A_0$ and $sign(\mu)$ one can find the
regions that are detectable by GLAST. 
Besides the previously explained parameters $m_0$, $m_{1/2}$,
$\tan\beta$ denotes the ratio of the vacuum expectation values of the
two neutral components of the SU(2) Higgs doublet, $A_0$ 
is the proportionality factor between the supersymmetry breaking trilinear
couplings and the Yukawa couplings while $\mu$ is determined (up to a
sign) by imposing the Electro-Weak Symmetry 
Breaking (EWSB) conditions at the weak scale.

\begin{figure}
    \caption{GLAST reach in the CMSSM parameter space 
    \label{tanbe55cmssm}}
        \includegraphics[scale=0.8]{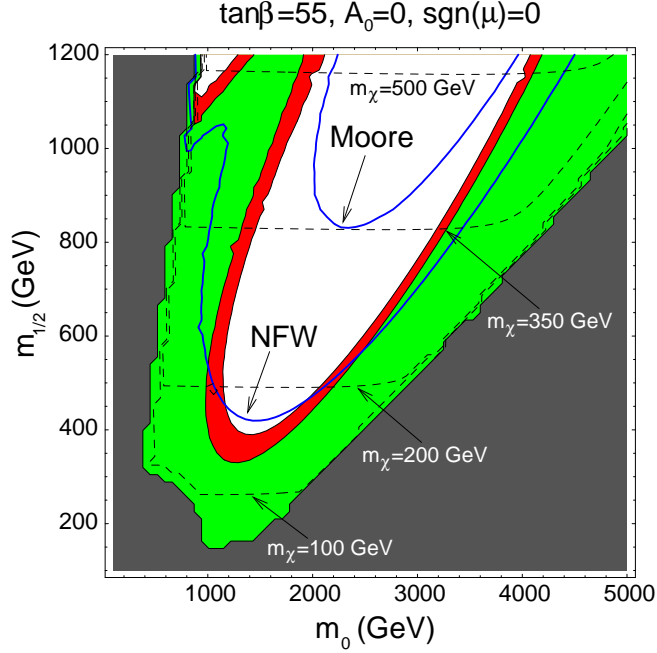}
\end{figure}

The result, at a $3\sigma$ confidence level, for $\tan\beta=55$
case is shown in fig.~\ref{tanbe55cmssm}. We assumed a total exposure of
$3.7\times 10^{10}\, {\rm cm}^2\, s$, an angular resolution (at $10$
GeV) of $\sim 3\times 10^{-5}$ sr and 4 years of data taking~\cite{glast-lat-perf}.  

Blue solid lines represent the GLAST reach for a NWF and Moore profile
while dashed lines represent the neutralino isomass contours
(expressed in GeV). Gray
shaded regions are parameter space regions excluded by either
theoretical or experimental (accelerator) constraints.
The upper left region is excluded because the lightest stau
is the LSP, the lower left region is excluded due the accelerator
bounds on the Higgs boson masses,
$b\to s \gamma$, slepton and squark masses, etc., while the right lower region is excluded because there is no
electroweak symmetry breaking.
The WMAP compatible region (with relic density $0.09\le\Omega h^2\le 0.13$) is
the red one while in the green region the
neutralino is a subdominant dark matter component (with $\Omega h^2\le 0.09$).
  
\section{Extra Dimensions and GLAST}
\label{extradim}
In recent times many scenarios involving extra space-time dimensions
with or even without supersymmetry received great attention. It is
very interesting to see if GLAST is able to detect signals in
this context. We consider a very wide scenario involving
extra dimensions and low energy supersymmetry~\cite{Fucito:2006ch}. 

In this framework there are three additional parameters, namely the
typical size of extra dimensions $\mu_0=R^{-1}$, their number $\delta$
and the number $\eta$ of fermion generations that are allowed to have extra
Kaluza-Klein (KK) states. The most striking feature in this class of models
is the presence of power-law corrections to all the couplings and
masses of the MSSM~\cite{Dienes:1998vg}. This implies that one can
have a grand unification scale (GUT) scale as low as few TeV.
From the point of view of indirect detection the case $\eta>0$ in which
some fermion generation have KK tower is disfavored. Thus in the following
we concentrate on the minimal case $\eta=0$. Moreover the number of
extra dimensions $\delta$ does not seem to be crucial for the
neutralino phenomenology~\cite{Fucito:2006ch}.
In almost all the parameter space the neutralino is still the LSP. One
of the main result is that, unlike the standard CMSSM case, the
neutralino is no longer bino-like but it tends to be a very pure
higgsino.
In general this result holds for not too high compactification scale
$\mu_0\lesssim 100$ TeV while going towards higher scale the
neutralino tends to be a bino.

We show the $3\sigma$ GLAST reach (along the line sketched in the
previous section)
for $\mu_0=10$ TeV, $\delta=2$ and $\eta=0$ and for a low value of
$\tan\beta$.

\begin{figure}
    \caption{GLAST reach for the extra dimensional scenario with
      $\mu_0=10$ TeV, $\delta=2$ and $\eta=0$ 
    \label{glast reach extradim 104}}
        \includegraphics[scale=0.8]{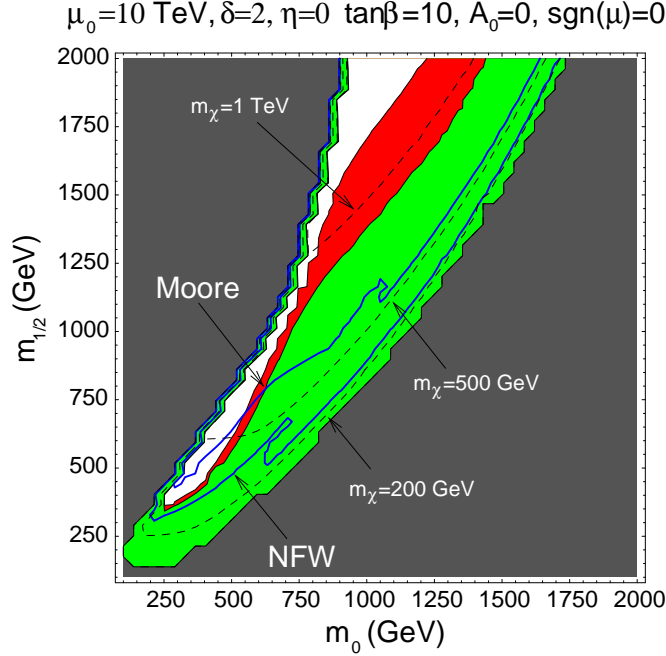}
\end{figure}

\begin{figure}
    \caption{GLAST reach for the extra dimensional scenario with
      $\mu_0=10^5$ TeV, $\delta=1$ and $\eta=0$ 
    \label{glast reach extradim 108}}
        \includegraphics[scale=0.8]{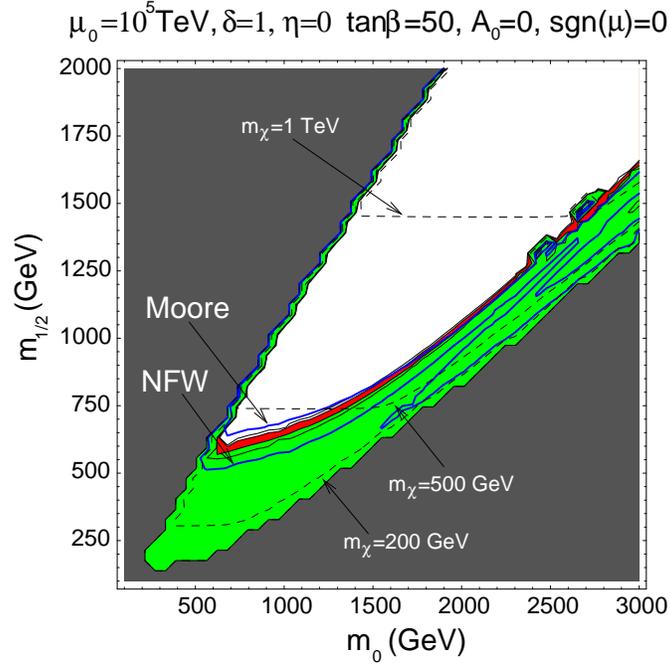}
\end{figure}

The plot assumptions and the explanation of the different regions are
the same as in fig.~\ref{tanbe55cmssm}. 
In this scenario a NFW profile is not enough for GLAST to be
able to explore the cosmologically favored region in which the
neutralino has a mass greater than $200$ GeV. In order to do that
a Moore profile is needed.
This corresponds to an enhancement factor of about one order of
magnitude in the continuum flux.  
It has
to be remarked that assuming some non standard cosmological evolution the
green region could become a WMAP compatible region. This is due to a non
thermal enhancement of the relic density (see for
example~\cite{Moroi:1999zb}). These kind of scenarios are particularly
motivated in scenarios with extra dimensions.
In fig.~\ref{glast reach extradim 108} we show the case with
$\mu_0=10^5$ TeV and with only one extra dimensions $\delta=1$. 
All the KK particles are quite heavy in
this case but the effects on the low energy theory, through one loop
processes, are still sizable.

In this scenario the GLAST reach for a NFW profile is still below the
WMAP allowed region but for a Moore profile GLAST will be able to
probe almost all the cosmologically favored region.

\section{Conclusions}
We have shown that, assuming a NFW profile, GLAST will be able to probe a huge part of the WMAP
allowed zone of the CMSSM parameter space especially in the high $\tan\beta$ case. 
In the scenario involving extra dimensions, with quite low
compactification scale $\mu_0\lesssim 10$ TeV, GLAST will still be able to
probe some part of the cosmologically favored region of the parameter
space either in the case of more cuspy Moore profile or in the case of
an enhancement of the neutralino relic density in a
non standard cosmological scenario.
In the case of higher compactification scale $\mu_0\gtrsim 10^5$ TeV
GLAST will be able to cover almost all the WMAP region.
\label{concl}

\end{document}